\documentclass[prd,preprint]{revtex4}

\usepackage{graphicx}
\usepackage{latexsym}
\usepackage[center]{subfigure}

\begin{document}
 \newcommand{\bq}{\begin{equation}}
 \newcommand{\eq}{\end{equation}}
 \newcommand{\bqn}{\begin{eqnarray}}
 \newcommand{\eqn}{\end{eqnarray}}
 \newcommand{\bqs}{\begin{equation}\begin{split}}
 \newcommand{\eqs}{\end{split}\end{equation}}
 \newcommand{\nb}{\nonumber}
 \newcommand{\lb}{\label}
 \newcommand{\p}{\partial}

\title{Thermodynamics of Time Machines}
\author{Michael Devin}

\begin{abstract}

In this note, a brief review of the consistent state approach to systems containing closed timelike curves[CTCs] or similar devices is given, and applied to the well known thermodynamic problem of Maxwell's demon. The 'third party paradox' for acausal systems is defined and applied to closed timelike curve censorship and black hole evaporation. Some traditional arguments for chronology protection are re-examined.\\\\
``... it's more like a big ball of wibbly-wobbly, timey-wimey... stuff.'' \\
- \emph{Doctor Who} 
\end{abstract}

\maketitle

Since the original version of this paper in 2001\cite{me}, there has been a renewed interest in time machine calculations, springing from a duality between acausal systems constrained by a quantum analog of the Novikov\cite{nov1} consistency principle, and the formalism of post-selected ensembles developed by Aharnov and many others\cite{seth1,ahr1,kip}. Interest has also grown in the applications of such systems to computation theory, following the footsteps of Deutsch\cite{deut}, who employed a different superselection criteria leading to different physics. The original aim of the paper was twofold. First, to develop the machinery for making calculations of acausal systems, and secondly, to apply the principles of thermodynamics to such systems. In the past ten years the body of work of post-selected ensembles has grown to become the standard concerning time machines\cite{seth1}, obviating much of the need for the first. Perhaps the second will find a better reception now. Some material has been added to reflect the more recent developments in black hole physics and post-selected systems.  

\section{Formalism}

Our time machine consists of an in state qubit and an out state qubit, related by some evolution operator. At the beginning of the experiment the qubit emerges in an unknown state $\psi_{out}$. We can perform a projection measurement of this onto some basis giving either $|+\rangle$, or $|-\rangle$. Later a state $\psi_{in}$ is sent into the time machine which acts on it to produce the original state. If the operator taking $\psi_{in}$ to $\psi_{out}$ is a unitary operator, say A, then we can produce a contradiction by choosing an in state of 
\bq
\psi_{in} = A^{-1}(|+\rangle\langle-| + |-\rangle\langle+|) P \psi_{out}
\eq
where P is our measurement projecting $\psi_{out}$ onto either $|+\rangle$ or $|-\rangle$. The probability of our measurement outcome would then be
\bq
p_+ = |\langle \psi_{out}|A|\psi_{in}\rangle|^2 = |\langle +|A A^{-1}\left( |+\rangle\langle -| + |-\rangle\langle +|\right)|+\rangle |^2 = |2Re\left[\langle +|-\rangle\right]|^2
\eq
thus giving a zero probability for our initial measurement. This would be the analog of the grandfather paradox. The problem is resolved by adding noise to the channel that connects the in states to the out states. If we take the noisy channel to have a bit error rate of $k$, then we may use that probability as relative weight to renormalize the possible histories of the system containing the time machine and arrive at predictions for measurements. Those histories in which the out state equals the in state gain a weighting of $1-k$, and those with $\langle\psi_{in}|\psi_{out}\rangle=0$ gain a weight of $k$. A more complete description would give the weight for each doublet of in/out states as a joint distribution, corresponding to a nonunitary, stochastic, internal evolution of the time machine.  For the time machine weight function $\omega$ we require,
\bq
\omega ( \psi , A\psi )  > 0
\eq
for all unitary $A$, for at least one $\psi$, to eliminate the grandfather paradox. The treatment is otherwise the same as a conventional sum over histories approach, but with the added weight term $\omega ( \psi_{in} , \psi_{out} )$ to the probability of each measurement outcome, renormalized. Classically the value of $k$ would be all that is needed to describe a single bit time machine, and an effective value of $k$ for the quantum time machine may be derived from the weight function. If no measurement is made of the state $\psi_{out}$, but it is acted on by some unitary rotation before re-entering as $\psi_{in}$, then an invariant state exists up to an overall phase. To prove this we simply apply the Brouwer fixed point theorem to the Bloch sphere representation of the qubit. We may consider the noise in this case to be an intrinsic uncertainty in the phase of the bit. For an initial measurement of $\psi_{out}=|+\rangle$, we may choose a unitary $A$ to minimize the ratio,
\bq
\frac{k}{1-k}=\frac{\int |\langle + | \phi \rangle|^2 \omega ( \langle\phi | , A | + \rangle ) d\phi}{\int |\langle + | \phi \rangle|^2 \omega ( \langle\phi | , | + \rangle ) d\phi}
\eq
to define the effective $k$ after integrating over all qubit states.
Consider a thought experiment where we take the out bit, measure it, then act on it with some unitary operator $A$, before sending it into the in channel. The probability of measuring $|+\rangle$ would be,
\bq
P(+) = z^{-1}(1-k)|\langle+|A|+\rangle|^2 + z^{-1}k|\langle-|A|+\rangle|^2,
\eq
with a normalization factor
\bqn
\nonumber z = (1-k)|\langle+|A|+\rangle|^2 + (1-k)|\langle-|A|-\rangle|^2 \\
+ k|\langle-|A|+\rangle|^2 + k|\langle+|A|-\rangle|^2.
\eqn
This factor corresponds to the four possible histories for the system, two for each possible result of the initial measurement. The two terms in the probability correspond to the case where $A$ acts on the bit, and the noise does not reverse it, and the other case where the noise offsets the rotation of the bit by the operator $A$. In each case $|+\rangle$ is still measured and sent to $A$. This scheme is formally identical with a post-selected system with two measurements. One an initial state preparation of $\psi_{out}$, and the other a post-selection measurement after both $A$ and the noise have acted on the bit. Each approach gives the same probabilities for the results of measurements on the system in between the in and out events. In the analogous post-selection experiment, without noise the sample size of ensembles meeting the post-selection criteria goes to zero as $\langle\psi|A|\psi\rangle$ does, leading to a crisis. With noise the denominator remains nonzero. Our second thought experiment will help give more justification for adding the noise.

\begin{figure}
\includegraphics[scale=.3]{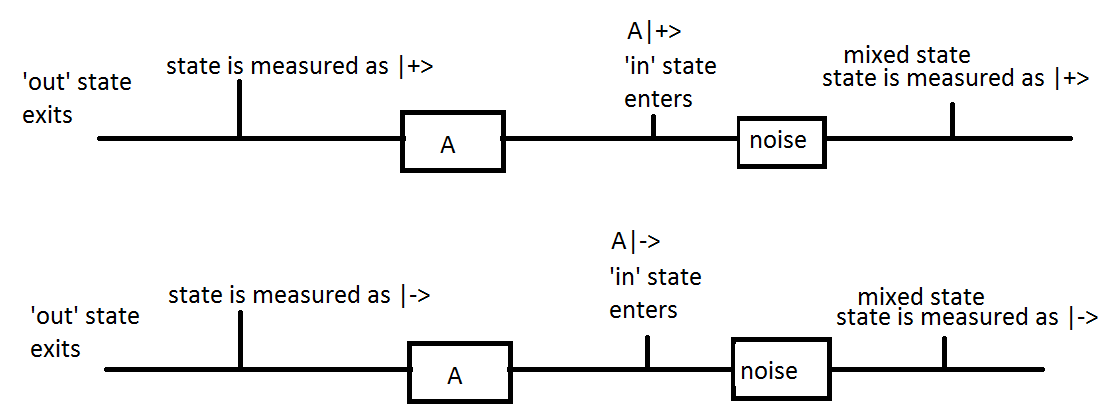}
\caption{Two principal outcomes of a generic single bit time machine}
\end{figure}

In the language of post-selection, this model assumes that rather than a random sample of a post-selected ensemble, we have a sample of a composite ensemble where a fraction $2k$ of ensembles with uncorrelated in and out states are admitted. Finally,
\bqn
\nonumber \rho_{out}(A) = (1-2k)\int \frac{\langle\phi|A|\phi\rangle^2}{Z_1(A)}|\phi\rangle\langle\phi|d\phi \\
+ 2k\int \frac{\langle\phi|A|\psi\rangle^2}{Z_2(A)}|\phi\rangle\langle\phi| d\phi d\psi
\eqn
where $Z_1,Z_2$ are the individual normalization factors for each integrated distribution respectively. It should come as no surprise that the emergent mixed state depends on the whole evolution outside the time machine $A$. Now $\rho_{in}$ will simply be unitarily related to $\rho_{out}$ by $A$. The first term is the equivalent of a normal post-selected ensemble where the state is required to be periodic and the initial and final measurements are evenly distributed over the space. The second is a similar even sampling over ordinary state  evolutions, that is, regular unitary quantum statistics evolving from a maximum entropy mixture.

\section{Heat and Work}

Suppose we take the time looping bit and entangle it with the position of a particle in a box. The box is divided by a partition into two unequal sections. In the case of a classic Szilard engine we measure which side of the box the particle in on and then adiabatically expand the section with the particle to the size of the original box, performing work. by Landauer's principle we generate one bit of entropy reseting the measurement apparatus, which is exactly equal to the maximum work we can extract from the engine by placing the partition in the center. When the particle is entangled with the time machine qubit, the probability distribution is no longer uniform and net work can be extracted.

Begin the thought experiment with a single particle in a box. A partition is inserted dividing the box into two unequal regions, of volume $x$, and $1-x$, respectively. The partition is moved sideways, to another location, $y$, performing some work. The expectation value of the work done is
\bq
W = p \ln{\frac{y}{x}} + (1-p)\ln{\frac{1-y}{1-x}}.
\eq
In this case $p$ is the probability of the particle being trapped in the region $x$. In the classical non-entangled case: $p=x$, but in this case we arrive at 
\bq
p = \frac{(1-k)x}{x(1-k)+k(1-x)}
\eq
by setting up the box as the control bit in a c-not gate acting upon the bit circulating in the time machine. Of the four histories of the two bit system, ones that have the particle in region $x$ are given a higher weight relative to ones that have the particle in the opposite chamber. The relative weight factor is
\bq
f = \frac{1-k}{k}.
\eq  
Maximizing the work function quickly gives 
\bq
y=p.
\eq
The work reduces to
\bq
W= p \ln{(\frac{1-k}{k})} + \ln{\frac{k}{k + (1-2k)x}},
\eq
where the ideal partition is placed at
\bq
x = \frac{k(1-k)}{(1-2k)^2}(\ln{\frac{1-k}{k}}) - \frac{k}{1-2k}
\eq
and is moved to
\bq
y = \frac{1-k}{1-2k} - \left( \ln{\frac{1-k}{k}}\right)^{-1}.
\eq
The final state of the partition is a uniform probability density that leaves the entropy of the box maximized. The maximized work as a function of only $k$ is
\bqn
\nonumber W = -\ln k -\ln{\ln{\frac{1-k}{k}}} -1 + \ln{(1-2k)}+ \frac{k}{1-2k}\ln{\frac{1-k}{k}}.
\eqn
This work function is dominated by the first term over the realistic domain for $0<k<1/2$. As the noise drops to zero for a time machine, the work extractable per bit diverges. A perfectly reliable time machine can therefore violate the second law of thermodynamics by an arbitrarily large amount, but a noisy one has an effective limit. Considering an assembly of $N$ such time machines acting in concert, we arrive at an effective value for $k_{eff.}=k^N$. If each time machine generates entropy of at least $-\ln k$, then the second law could be preserved. The second term with double logarithm is interesting as well, reminding one of the postulated logarithmic corrections to black hole entropy proposed by some theorists.

\begin{figure}
\includegraphics[scale=.3]{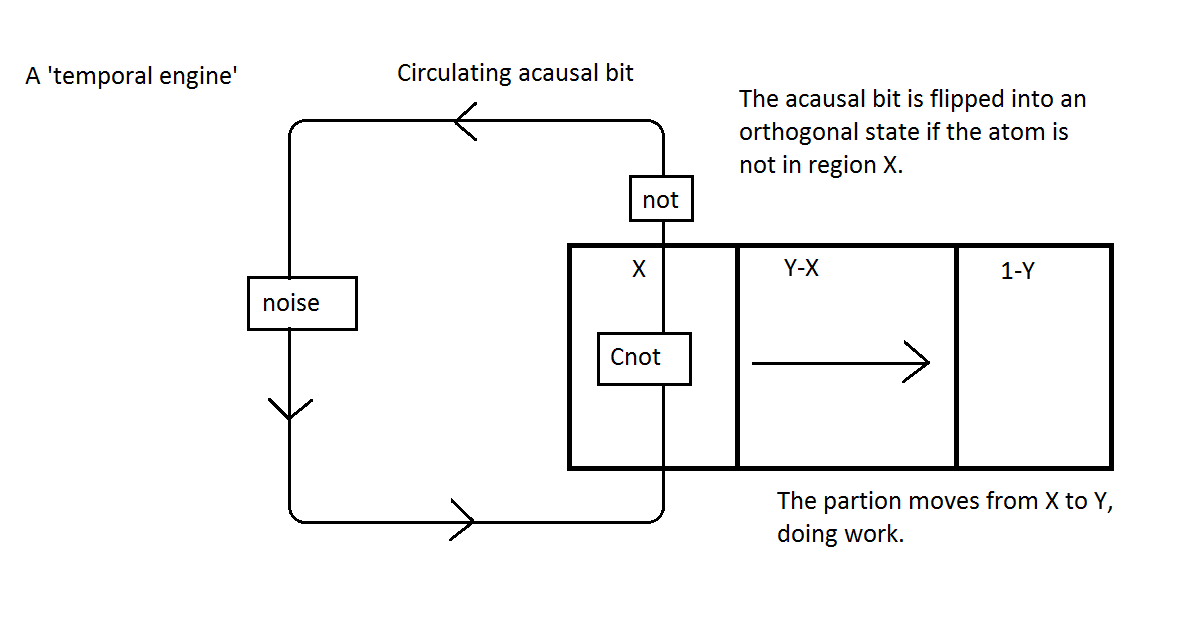}
\caption{Time loop acting as Maxwell's demon}
\end{figure}

Internally the time machine qubit should have von Neumann entropy of 
\bq
S_{int.} = p \ln{p} + (1-p)\ln{(1-p)}.
\eq
Whether this entropy should be considered a cost of establishing the time loop is unknown, and it was not considered in the work maximizing calculations. Likewise, other protocalls taking into account measurements made on both the time machine bit and on the atom, may give differing work functions. In the case where the atom is observed and the piston moved conditionally, so that work done is always positive, but a bit must be erased afterwards, the work maximization is not expressible in terms of elementary functions, but is approximately described by W-Lambert functions. It is tempting to cite Landauer's principle and conclude other schemes that add measurement do not on average add net work when erasure is taken into account.

Considering machines with larger numbers of internal states one concern that arises is temperature. A normal thermodynamic system is constrained by conservation of energy to borrow energy from it's environment. In the classic derivation of the canonical ensemble the number of microstates available to the environment with that borrowed energy produces the well known Boltzmann factor. Systems with positive finite temperature and finite degrees of freedom have finite energy. In some of the cases considered in general relativity, with back reactions ignored, we find that CTCs and other time machines act like systems that do not need to borrow from anywhere to have energy. The number of accessible states grows exponentially with energy, and with all microstates equally probable, we naturally arrive at a negative temperature. A similar argument may be used for particle number to give negative chemical potential to the occupation numbers of each field mode. If the number of particles or energy is not somehow bounded then a divergence can result. This is especially the case when we have the internal entropy naturally maximized by eliminating the interaction of the time machine with it's environment due to ignoring back reaction. 

The appearance of these divergences is often cited as support for Hawking's chronology protection conjecture\cite{hawk6,cass1}. It is assumed that the fluctuations must destroy the time machine before anything improper can occur. However, if this is the case, then it provides the very mechanism for making time machines well behaved entities with positive temperature. The higher the energy or occupation number of a particular field mode in a time machine, the more it is suppressed by the re-weighting of histories by the amplitude for such a high energy state to scatter onto the same state. In post-selected language the sample of high energy states acceptable to post selection is small because high energy modes tend to decay, and high particle number states tend to dissipate, with exponential probability. 

For example, if N bottom quarks will probably not come out of a wormhole because, given N such quarks emerge, it would be unlikely that they all fail to decay and somehow scatter back into the wormhole later on, especially if the time lag was large between mouths. The weighting of internal states in these cases is not dominated by noise, but rather by the much smaller probabilities for conventional scattering processes outside the time machine to result in a naturally periodic state. Time machine temperature will likely drop dramatically as the lag time between in and out states increases, or as the energy approaches an effective maximum given by the time machine's capability to withstand the back reaction effects. 

Hawking also calculates the entropy of a scalar field in a rotating spacetime that is gradually shifted into an acausal spacetime by adding angular momentum to the field\cite{cass1}. He finds that the entropy diverges to negative infinity, which would imply zero allowed states and thus perhaps a `quantum prohibition' of the transformation. This should be a generic result for fields on classical backgrounds, since they contain no source of noise, and are thus described by setting $k=0$. The system is capable of extracting arbitrarily large amounts of work from an entangled system. In general we can imagine that systems with very large values of time machine negentropy will behave quite strangely, as the probability of exotic events could be exponentially amplified. It would be reasonable to assume that quantum gravity or other small scale physics will become relevant at much lower energies than they normally would. As far as wormholes and acausal GR spacetimes go, a geometric uncertainty relation could provide the necessary noise work tradeoff to preserve good thermodynamic behavior. Consider a bit looped through a wormhole and then phase flipped. If the phase contributed by the wormhole obeys an uncertainty principle similar to the normal Heisenberg time-energy relationship, then the effective $k$ could yield a work function of less than the energy needed for such accuracy.

In regards to the computational power of computers with access to time machines, it is straightforward to see that any computation in which an efficient method for checking the solution exists, a source of random bits, sent through a checking algorithm which then acts as a c-not on a time machine qubit, can arrive at the correct solution immediately if the informational content of the solution is sufficiently less than the work function of the time machine. Since time machine bits may also act as perfectly random sources, the information may seem to be created from nothing, but one may also think of such `calculations' as becoming an extremely lucky guesser, due to the re-weighting of histories by the time machine. 

Essentially time machines are entropy pumps, similar to a classical heat engine. Instead of transporting heat, they transport entropy, pushing a system of particles or a random message in message space into a lower entropy state, but increasing the entropy of the environment in some way not yet understood. The computations, like those of a true quantum computer, are essentially analog computations. In this case effectively physically reversing classical hard problems in time. Conventional crytpography would pose little obstacle to such a computer. Instead one would have to devise ambiguous codes, which could be decoded into a number of intelligible but incorrect messages, leaving the computer with a problem of interpreting which was significant, a task made hard for entirely different reasons. A `brute force' entropy attack assisted by a time machine would then more likely generate one of the red-herring messages. Other unusual protocalls might be used to increase security, but public key certification by computer would be almost useless.

\section{Unitarity}

To new students of quantum mechanics, the Bell inequalities, delayed choice, and quantum eraser experiments have seemed to almost violate causality. The fact that they cannot is a crucial consequence of the unitary nature of quantum mechanics. One of the most troubling aspects of the information loss paradox is the apparent loss of unitarity. Not all non-unitary maps are created equal, and trace over models of lossy processes do generally preserve causality. Such models seemed adequate until Hawking radiation came along. The eventual disintegration of the hole broke the analogy of environmental decoherence opening up the possibility of `bad' nonunitary processes in some imagined acausal lossy theory of quantum gravity. The aim of the remaining sections is to explore implications of this possibility.

A quantum eraser is a system that exhibits extreme nature of the delayed choice experiment by measuring and then coherently erasing information about two different possible paths for a system. By the no copy theorem a qubit that is created by measuring another qubit can only be coherently erased by combining it with the original again. Coherent erasure makes the erased bit `unrecoverable in principle' and thus restores interference effects relating to any superposition of the original bit before the creation of the measurement bit. Two concerns in the information paradox were first, that an evaporated black hole might constitute an `in principle unrecoverable' process, and second that proposed complementarity scenarios would violate the no copy theorem, providing another way to erase measurements. Both cases lead to breakdown of unitarity and subsequently causality. Complementarity has to ensure the second scenario of a bit meeting its extra twin can not occur. This appears to be the primary motivation for the recent 'firewall' models of black hole evaporation.

\begin{figure}
\includegraphics[scale=.3]{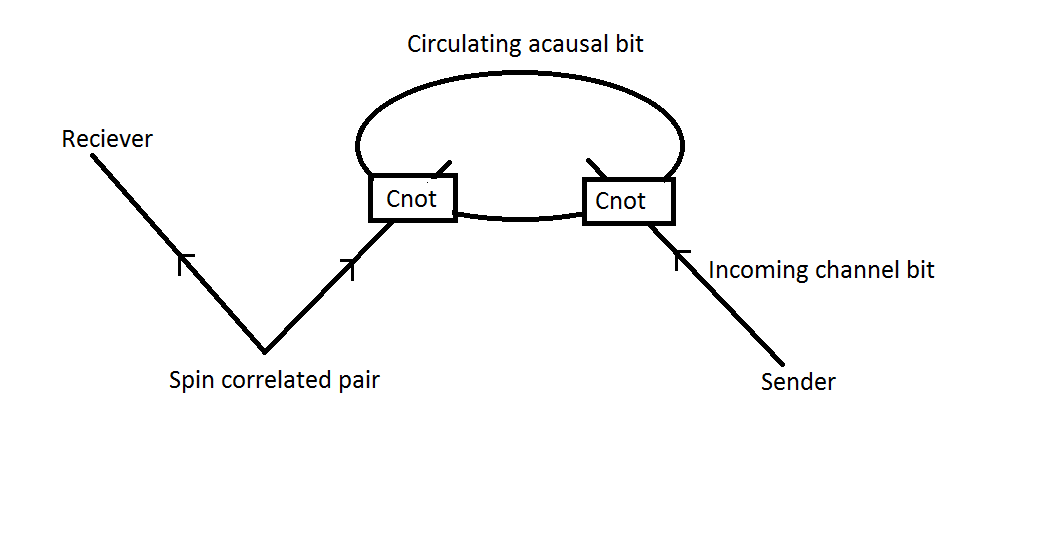}
\caption{Third party pardox communication channel}
\end{figure}

The inherent non-unitarity of time machines can easily be seen by observing the effect that this probability skewing has on entangled particle pairs. Consider instead of a particle in a box, the classic spin entangled pairs of particles. If we should choose one of the entangled particles to be sent an arbitrary distance away, then use the other as a control bit in our time machine circuit, then the state of the pair becomes in general a mixed state. If we designate a second communication channel to act as the control of another c-not gate on the time machine bit, then we may measure a real correlation between that channel and the spin measurements of the distant spin partner. A single time machine as a third party in the mutual future of two observers can apparently effect nonlocal communication between them. Thus the non-unitary effects of a time machine may be felt arbitrarily far away, even in the past light cone of the arrival of $\psi_{out}$. 

Consider the equivalent case for a post-selected system where a single bit is selected to be in a state $|0\rangle$ at two different times. In between these two it is acted on by by two controlled not operations, one of an EPR pair, and a second being the desired input bit. The post-selected statistics of the distant EPR partner will now reflect those of the chosen input bit. Any time a superselection operator acts on an entangled system of particles to enforce a particular final state on part of the system, the potential for acausal communications between two third parties also appears. This `third party paradox', is an important element in understanding the interaction between time machines and nonunitary dynamics. 

So far it seems that time machines skew the statistics of ensembles to create effective nonlinear dynamics. In turn most nonlinear quantum mechanics appears to be exploitable to create time machines. Explicitly, one time machine can be used to create another, or any number of others, through the third party paradox. A useful exercise here is to consider the work done by these `child' machines and how it compares to the work extractable by the parent alone. Each child `inherits' the noise of its parent, and shares to some degree the back reaction of its siblings. If the spawning process introduces no additional noise, then we can shift the arrival time of $|\psi_{out}\rangle$ to an earlier time and find an equivalent system containing only the parent time loop. This is possible since the duration of the loop is not a factor in the work function. The maximum work performed by the entire ensemble, minus any entropy cost for erasing extra measurement bits, should still be less than or equal to the original work function.

 Early in the `black hole wars' Hawking tentatively proposed a theory of density matricis might be considered as a generalization of quantum mechanics capable of handling the apparent lack of unitary evolution in gravitational collapse\cite{hawk5}.  This approach was heavily criticized for possible violations of locality or energy conservation\cite{sus1}. Energy conservation can be maintained, but the trade off between causality and non-unitarity remains. Any system that can act on a qubit to map orthogonal to non-orthogonal states, can be added to a quantum eraser double interferometer to break the fine balance between opposing interference patterns that locally cancel for distant entangled states. It would seem though that if such transitions were possible, then vacuum fluctuations would cause them to occur to any given bit eventually, and thus nonlocal interactions would be everywhere.

The proliferation of such communication channels is hampered somewhat by the fact that in the case of communication through a nonunitary map, it's quite possible that the phase difference between the erased states will also affect the interference at the receiver. If the map is sufficiently far in the future, or the record bit is further entangled before being erased, this could give the signal a large noise component proportional to the uncertainty in the phase of the erased bit. Presumably virtual interactions in any nonunitary extension would occur significantly only at very high temperature, and would thus be swamped by conventional unitary decoherence effects long beforehand.

Hawking and others have contended that all systems containing time machines should possess entropy in accord with the number of internal states `sent' to the past\cite{hawk4}. This effect can be seen from the following thought experiment. We take a single qubit time machine and use the $\psi_{out}$ bit as the control bit acting on an incoming pure state $|+\rangle$. $\psi_{out}$ is then returned into the time machine as $\psi_{in}$, leaving behind the acted upon qubit. The value of this bit is apparently not determined by any states outside the system, and both values correspond to equally weighted histories. The natural conclusion is that the originally pure state $|+\rangle$ has been mapped onto a maximally mixed qubit. Entropy has increased, although only by $\ln 2$. This scenario is trivially modeled in a post-selection experiment as simply three measurements of a random bit, in which the first and last measurements are the same result. 

\begin{figure}
\includegraphics[scale=.4]{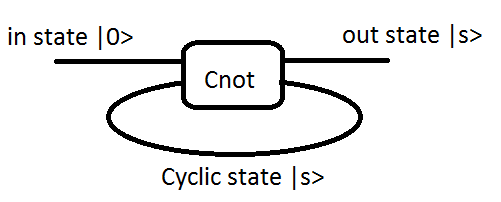}
\caption{Hawking's circulating states and the unproven proof paradox}
\end{figure}

A similar case of tracing was thought to occur in black hole physics not just over collapsed matter but over those modes of Hawking radiation that both begin and end on singularities or similar boundary surfaces. Two examples would be a double black hole system with radiation moving between one hole and another, and a single black hole surrounded by a reflective shell, that would redirect radiation back into the hole. These cases appear to be resolved along with the information paradox itself with the rise of the string theory model of the horizon and lossless black holes. They remain important problems for any lossy theory of quantum gravity, however.

\section{Black holes}

If time machines are possible, they should be accompanied by entropy producing phenomena. The most reasonable candidate would seem to be the black hole and its associated information paradox. If black hole formation and evaporation is nonunitary, then this opens up the possibility for time machines to be constructed out of the resulting unitary violations. One reason for this is the importance of the relative phase information of out states that is crucial to preventing entangled particle pairs from allowing non-local communication. The classic double interferometer fails to detect any local interference effects when observing only one of the photons. The other photon may be in either of two states, and that bit contains either the path information of its cousin, eliminating the interference, or the two outcomes contribute to two separate interference patterns exactly out of phase, such that the trace over those gives no local interference.

Some black holes are thought to contain ready-made time machines in the form of closed timelike curves. The troubling behavior of the Kerr metric near the singularity was assumed to be made safe by being behind the horizon, an early and important result supporting the cosmic censorship hypothesis. However due to the third party effect, it would appear that not only does the horizon fail to prevent information from leaving the CTC region, it leads non-local communication between points far from the hole. These secondary time machines can then effectively 'mine' the black hole for negentropy. Some fraction of the entropy associated with the irreducible mass of the hole should then provide a bound on this entropy, and therefore some constraints on $k$, for the CTC region. For the purposes of chronology protection, horizons alone are  ineffective 'transparent censors'.

The simplest case may be the most pathological. Consider if all incoming matter states are mapped to a single maximally mixed state, allowing perfectly reliable nonunitary erasure of quantum information falling into the hole. In that case a black hole assisted quantum eraser would be able to create a noise free acausal channel, thus allowing arbitrarily large second law violations. If at least some information escapes, either to the outside as Hawking radiation, into another universe, or frozen onto the singularity, then it must be traced over, and the eraser will gain noise and a corresponding negentropy bound. Comparing this bound to the standard black hole entropy produces a constraint on possible nonunitary quantum gravity processes compatible with the second law. The most well behaved causally are remnant and baby universe theories, as well as other models that do away with the black hole entirely, such as fuzzballs, but these are outside the scope of this paper.

One proposal in resolution to the black hole information paradox is to add a boundary condition to the singularity\cite{mal1}. Some critics argue this violates causality\cite{pres2}. The argument against it can be illustrated with the following paradox. Under normal circumstances, information, such as a volume of Shakespeare, falls into a black hole, which then evaporates via Hawking radiation. If a boundary condition at the singularity is prescribed, then these fields must be canceled by other contributions as they approach the singularity. These other contributions are the in falling components of the pairs of particles appearing from the vacuum, the outgoing of which constitute the Hawking radiation. Since each pair is strongly entangled, and the in falling radiation is forced to match up with the in falling Shakespeare volume via the superselection of global field configurations to fit the boundary condition, then the outgoing radiation must be simply the scrambled volume of Shakespeare. Another way of considering it is to imagine the field modes reflect off of the singularity, becoming negative energy time reversed modes. They then travel out of the hole and reflect off of the potential outside the main black hole, becoming again positive energy, forward modes.

The boundary condition acts as a selector of global field configurations, much like the post-selection operator used to model acausal ensembles. The proposed mechanism `similar to state teleportation' is in fact the third party paradox communication channel arising in both time machine and post selected systems. We may employ the same methods of superselection to generate a time machine via the third party problem. The picture is complicated slightly though by the presence of the incoming part of the Hawking pairs. This incoming part may serve as the required noise that bounds the total work extractable by all third party time machines. If no time machines are spawned this way, the work is expended adjusting the outgoing radiation into the form of Shakespeare. One flaw in this method of teleportation is also that there is nothing to require that the teleported states leave the black hole before the original states enter it.

In order to harness this boundary condition to create third party time machines imagine that we have an identical copy of the volume that formed the hole, call it anti-Shakespeare, that we created along with the original, so that we know all of the information that was lost in the initial process of collapse. Now, a the contrarian in possession of anti-Shakespeare decides to predicate his release of it into the black hole upon the decay of an atom, spin pair or other similar event. The same mechanism that forces the radiation to match the original Shakespeare will also affect the decay of said atom. If we can toss the anti-Shakespeare after the original in such a way as to quantum erase the original, a sort of pair annihilation, then neither will contribute to the deficit in field values between incoming matter and the prescribed boundary state. Since the black hole is nonrotating, it's mass will not be decreased, but instead increased by the mass of anti-Shakespeare. With a higher mass, but no change to the boundary conditions at the singularity, the incoming Hawking states now need only balance each other and whatever other mass besides our pair of books created the hole. The evaporation process now has exponentially more ways to `balance the books' when anti-Shakespeare is dropped than when it doesn't. Thus anti-Shakespeare will be prescribed to be dropped with probability of approximately 
\bq
p=1 - e^{-S_{as}}.
\eq
If we then use this effect to produce a third party communication channel, it will have a bit error rate of approximatey
\bq
k = 1-p
\eq
and thus a time machine work function of approximately
\bq
W = -\ln{k} = -\ln{e^{-S}}=S_{as}.
\eq
The extra entropy produced by the decorrelation of the outgoing radiation with the volume of Shakespeare, is dominated by the same term as the maximum work extractable from the resulting causality violation.  In that case the double interferometer setup would produce an error rate inversely proportional to the number of black hole microstates, allowing time machine work to balance Hawking radiation entropy again. A simple backwards calculation from $S=\ln 2$ suggests a possible lower bound on the error rate of about $7\%$, but other factors could easily alter this number. Using the duality with post-selected ensembles should allow us to do an experimental check on the Maldacena mechanism\cite{mal1}, as well as test for possible anti-Shakespeare-like phenomena. A more general model of a nonunitary black hole may contain some degree of information escape as well as loss. These hybrid models would be constrained by the extractable work and the second law. Of course the example here should be taken with a grain of salt. Interaction with Hawking pairs inside the black hole has been neglected, and depending on the nature of quantum gravity, anti-Shakespeare might be required to have negative energy or some other unphysical property, since it is a `reflection' of the original about the final state. 

\section{Tourism}  

Hawking famously has cited the lack of future tourists as good evidence against time machines. Although no one disputes this, it is an interesting case to consider for the would be time theorist. One possible explanation for the lack of such `tourist' fields on the microscopic scale could be something like the quantum zeno effect. The atom is locked in its state and the cat never dies because we generally have good records of whether or not time travelers have appeared. For such a traveler, our present would be his past, and such records in that future of a lack of visitors from the future may act as a similar lock on using tunneling or entanglement type phenomena as time machines for that purpose. Different possible tourists may destructively interfere with each other, just as highly non-classical paths for macroscopic systems do in path integral theory. Consider that the weight of a particular time tourist scenario is determined by the amplitude for the tourist's state to scatter onto itself at his later departure. For any large number of bits of the tourist, as those bits decohere with the environment that weight should decrease exponentially. 

A physical example of how one might look for such `tourists' could be realized by exploring the third party paradox where the receiving channel is measured well before the time machine exists. The spin measurements of that channel should be random, but if tourism is allowed, then they may contain a message. If we consider ensembles that may or may not contain a time machine, it is helpful to note that the weight factor for a particular history is an inner product of two unit vectors, as well as a noise coefficient. Both of these factors are less than one, and a sampling from ensembles where the existence of a later time machine depends on the reception of a message that enables it's construction will actually be suppressed relative to other random possible messages. A statistical `weak censorship' counteracts the spontaneous emergence of time machines, without absolutely forbidding them.  It might make for an interesting experiment to construct a post-selection equivalent of the tourist problem, in which selection criteria followed more complex protocols. 

In order for tourists to be sufficiently rare, the chronology protection mechanism need not be absolute. Instead it need only be exponentially difficult for tourists to visit some location in order for the expectation value of tourists to be finite, and thus hopefully small.

\section{conclusion}

In conclusion, time machines, if they exist at all, must possess fundamental limits on their error rate and waste heat, irrespective of the exact method of construction. These limits can be thought of as analogous to the old Carnot efficiency of classical heat engines independent of the specific construction of the engine. Most of the standard paradoxes associated with time travel are mitigated by considering systems operating within these limits. The study of acausal models still has much room for development. In the case of renormalization, badly behaved bare models may form condensates, shifting the vacuum and creating a more well behaved dressed model. Similarly, acausal bare models may lead to better behaved approximately causal models when various corrections are accounted for. In cosmology and landscape theory, some physicists have sought a model for the emergence of the Lorentzian signature of the metric, a spontaneous symmetry breaking that creates time itself. If such ambitions are ever to succeed they surely have to entertain causality as potentially only approximate in the resulting cosmos.

\end{document}